\documentclass[showpacs,preprintnumbers,amsmath,amssymb,twocolumn]{revtex4}
\usepackage{graphicx,color}
\usepackage{bm}
\usepackage{color}

\usepackage[usenames]{xcolor}
\usepackage{subfigure}

\xdefinecolor{mylinkcolor}{rgb}{0,0,0.5}
\usepackage[
 	linktocpage,
	bookmarksnumbered, bookmarksopen, bookmarksopenlevel=2,
	breaklinks=true,
	colorlinks=true, filecolor=mylinkcolor, citecolor=mylinkcolor,
	linkcolor=mylinkcolor, urlcolor=mylinkcolor, menucolor=mylinkcolor,
]{hyperref}

\newcommand{\be}{\begin{equation}}
\newcommand{\ee}{\end{equation}}
\newcommand{\bea}{\begin{eqnarray}}
\newcommand{\eea}{\end{eqnarray}}
\newcommand{\nn}{\nonumber}

\renewcommand{\H}{{\cal H}}

\begin{document}

\hypersetup{
	pdftitle={Dirac equation in 5-dimensional spherically symmetric space-times},
	pdfauthor={Yves Brihaye, T\'erence Delsate, Nobuyuki Sawado, Hiroaki Yoshii}
}

\title{Dirac equation for sphercially symmetric $AdS_5$ space-time and application to a boson star in EGB gravity}
\author{Yves Brihaye$^a$}
\email{yves.brihaye(at)ummons.ac.be}
\author{T\'{e}rence Delsate$^a$}
\email{terence.delsate(at)umons.ac.be}
\author{Nobuyuki Sawado$^b$}
\email{sawado(at)ph.noda.tus.ac.jp}
\author{Hiroaki Yoshii$^b$}

\affiliation{$^a$Theoretical and Mathematical Physics Dpt.,
Universit\'{e} de Mons -\\ UMons, 20, Place du Parc, 7000 Mons, Belgium \\
$^b$Department of Physics, Faculty of Science and Technology,
Tokyo University of Science, Noda, Chiba 278-8510, Japan}

\date{\today}

\begin{abstract}
We discuss the Dirac equation in a curved 5-dimensional spherically symmetric 
space-time. The angular part of the solutions is thoroughly studied, in a 
formulation suited for extending to rotating space-times with equal angular momenta. 
It has a symmetry $SU(2)\times U(1)$ and is implemented by the Wigner functions. 
The radial part forms a Dirac-Schr\"odinger type equation, and existence of 
the analytical solutions of the massless and the massive modes is confirmed. 
The solutions are described by the Jacobi polynomials. 
Also, the spinor of the both large and small components is obtained numerically. 
As a direct application of our formulation, we evaluate the 
spectrum of the Dirac fermion in Einstein-Gauss-Bonnet space-time and the space-time of a boson 
star.  
\end{abstract}
\pacs{}
\keywords{}

\maketitle 
\tableofcontents

\section{Introduction}


We study the Dirac equation in the background of a curved $5$-dimensional
space-time where the angular part is expressed as a $S^1$ bundle over $CP^1$.
This form arises naturally in the case of the rotating space-time with equal angular
momenta, for instance for the Myers-Perry black hole \cite{Myers:1986un}.

While the Dirac equation in 5-dimensional Myers-Perry with single angular
momentum background has been studied in the literature (see
for instance \cite{Ida:2002ez, Casals:2006xp, Flachi:2008yb}), we have not 
find a detailed discussion about the angular harmonic with a manifest $SU(2)\times
U(1)$ symmetry. Note that in the case mentioned above, the angular part is
expressed as spheroidal harmonics. In this paper, we provide a detailed
discussion on the representation of the angular sector of the Dirac field with such
a manifest symmetry. We will focus on the problems of the non-rotating, since the
computations are much simpler and the angular basis has the same symmetries than
the isometry group of the rotating case with equal angular momenta. This work
will serve as a basis for treating the rotating case, which is currently 
under consideration.

In the case of equal angular momenta, the space-time is cohomogeneous one,
contrary to the case of a single angular momentum, which in fact has more in
common with the four dimensional Kerr space-time. The angular separation of
bosonic fields in such a background relies on the enhancement of the symmetry of the case of
the equal angular momenta and has been addressed in
\cite{Murata:2007gv}. In this work it was shown that the angular harmonics are
expressed in terms of Wigner
functions (see also \cite{Rocha:2014gza} for an application). In this paper, we
 extend this result to the Dirac case, again, starting with the simpler limit of
no rotation.

We construct the spinorial harmonics in this background and derive a
Schr\"odinger like equation in the static case. As illustrative applications, we
construct symmetric plane wave solutions in flat space-time. We further construct
normal normalizable and regular modes in both massive and massless $AdS$ cases. 
These normal modes are obtained algebraically. This was first considered in
\cite{Cotaescu:2007xv} for the massive case, and we do recover the same
spectrum. 

Let us state some non-exhaustive previously known results. The separability of
the Dirac equation in the 5-dimensional Myers-Perry background has been discussed in
\cite{Wu:2008df}. In this case, the separability is proved in the less
symmetric case, namely black holes with unequal angular momenta. Massless and
massive Dirac fields in $D$-dimensional deSitter space-time are studied in
\cite{LopezOrtega:2007sr}. While an analytic continuation of the spectrum from
deSitter to Anti-deSitter would in principle solve the $AdS$ case, it is not
the case. We discuss that in Section VII.

Reference \cite{Camporesi:1995fb} treats massless Dirac fields on a compact
$S^N$ space or hyperbolic $H^N$ space, $N$ being the number of dimensions. Note
that the case we deal with is different here since first we include a mass and
second we have an additional coordinate.

Here, we explicitly build an angular basis with the suitable symmetry for the
case of equal angular momenta. The fact that the space-time is separable in this
case is clearly not a big deal, here our goal is to present the technical
details of our derivation. 
The new result in this paper consists in constructing the angular basis in
terms of Wigner function, completing the analysis of \cite{Murata:2007gv}.

This paper serves as a basis for future
extension, such as coupling to non-vacuum background solutions, e.g. those
constructed in \cite{Brihaye:2013hx}, rotating background solution (with equal
angular momenta), quasinormal mode spectrum, and so on.
Our analysis applies to asymptotically flat as well as asymptotically $AdS$ or
$dS$ space-times and potentially has many application in $AdS/CFT$ context, brane
physics, higher dimensional black holes, for instance.

This paper is structured as follows: we first briefly review the background
metric with equal angular momenta, before reducing it to the static case. We
then discuss the symmetries of the angular sector. In Sec.III we present in
details a construction of the vielbein and $\gamma$ matrices in this
space-time. 
In Sec.IV the Dirac equation itself is presented and a Schr\"odinger like equation
for the spinor is derived. We discuss the angular variable separation, in
particular we construct explicitly the spinorial harmonics in terms of Wigner
functions. The parity of the basis is also discussed. 
In Sec.V, the formulation of the radial part of the Dirac-Schr\"odinger equation is given.   
Sec.VI is devoted to the complete, orthonormal plane wave basis in flat space-time. 
In Sec.VII, we construct analytical massless and massive Dirac spinor solutions 
in background $AdS$ vacuum. We also present a numerical method and solutions of 
the massive equations. 
In Sec.VIII, as the simplest application of our formulation, we examine the spectrum 
of the fermion coupled with a boson star.  
Finally, Sec.IX contains some concluding remarks.

\section{Background space-time}
\subsection{Metric and angular isometry group}
Let us consider a 5-dimensional background space-time of the form
\begin{eqnarray}
ds^2&=&-b(r)dt^2+\frac{1}{f(r)}dr^2+g(r)d\tilde{\theta}^2 \nonumber \\
&+&h(r)\sin^2\tilde{\theta} (d\varphi_1-\omega(r)dt)^2 \nonumber \\
&+&h(r)\cos^2\tilde{\theta} (d\varphi_2-\omega(r)dt)^2 \nonumber \\
&+&(g(r)-h(r))\sin^2\tilde{\theta}\cos^2\tilde{\theta} (d\varphi_1-d\varphi_2)^2
\label{metric}
\end{eqnarray}
where $\tilde{\theta}$ runs from $0$ to $\pi/2$ while $\varphi_1,\varphi_2$ have
the range from $0$ to $2\pi$.
Introducing new coordinates 
$$\theta=2\tilde{\theta},\varphi=\varphi_2-\varphi_1,
\psi=\varphi_1+\varphi_2,$$
the line element then reduces to
\begin{eqnarray}
ds^2&=&-b(r)dt^2+\frac{1}{f(r)}dr^2 
+\frac{g(r)}{4}(d\theta^2+\sin^2\theta d\varphi^2)
\nonumber \\
&&+\frac{h(r)}{4}(d\psi+\cos\theta d\varphi-2\omega dt)^2.
\label{metric2}
\end{eqnarray}
This form of the space-time is well suited to study the spectrum of the Dirac
equation in several contexts such as those considered in \cite{Brihaye:2011fj}. In
fact, in the present paper, we will first consider
the static case, i.e. no rotations, but keeping this representation of the
angular sector, where the symmetries in the case of rotation with equal angular
momenta is more manifest.
Note that this form of the metric also accommodates black holes space-time, such
as the 5-dimensional Myers-Perry black hole metric in both asymptotically
$AdS$ and flat spaces.

In \cite{Murata:2007gv}, the authors extensively studied separability of field
equations in the degenerate Myers-Perry black hole metric (\ref{metric}).
They successfully decompose scalar, vector and tensor fields on the metric and
then we employ the method to a spinor fields on the static case. 

The static form of the metric (i.e. with $g=h,\ w=0$) has symmetry of
$SO(4)\simeq SU(2)_L\otimes SU(2)_R$ rotation group, which breaks to
$SU(2)_R\times U(1)_L$ when the rotation is present. 
We define two invariant one-forms $\sigma_a^{R,L}(a=1,2,3)$ of $SU(2)$ which
satisfy $d\sigma_a^R=1/2\epsilon^{abc}\sigma_b^R\wedge\sigma_c^R$ and 
$d\sigma_a^L=-1/2\epsilon^{abc}\sigma_b^L\wedge\sigma_c^L$. The explicit forms
are
\begin{align}
&\sigma_1^R=-\sin\psi d\theta+\cos\psi\sin\theta d\varphi\notag\\
&\sigma_2^R=\cos\psi d\theta+\sin\psi\sin\theta d\varphi\notag\\
&\sigma_3^R=d\psi+\cos\theta d\varphi
\end{align}
and 
\begin{align}
&\sigma_1^L=\sin\varphi d\theta-\cos\varphi\sin\theta d\psi\notag\\
&\sigma_2^L=\cos\varphi d\theta+\sin\varphi\sin\theta d\psi\notag\\
&\sigma_3^L=d\varphi+\cos\theta d\psi.
\end{align}

The metric (\ref{metric2}) is given by
\begin{eqnarray}
ds^2&=&-f(r)a^2(r)dt^2+\frac{1}{f(r)}dr^2 \nonumber \\
&+&\frac{g(r)}{4}((\sigma_1^{R,L})^2+(\sigma_2^{R,L})^2+(\sigma_3^{R,L})^2)
\nonumber \\
&-&\frac{g(r)}{4}(\sigma_3^R)^2+\frac{h(r)}{4}(\sigma_3^R-2\omega dt)^2.
\label{metric3}
\end{eqnarray}

In the static case, the metric has the the following Killing
vector
$\xi_{\alpha}~~(\alpha=1,2,3)$:
\begin{align}
&\xi^R_1=-\sin\psi\partial_{\theta}+\frac{\cos\psi}{\sin\theta}\partial_{\varphi
}-\cot\theta\cos\psi\partial_{\psi}\notag\\
&\xi^R_2=\cos\psi\partial_{\theta}+\frac{\sin\psi}{\sin\theta}\partial_{\varphi}
-\cot\theta\sin\psi\partial_{\psi}\notag\\
&\xi^R_3=\partial_{\psi}
\end{align}
and
\begin{align}
&\xi^L_1=\sin\varphi\partial_{\theta}-\frac{\cos\varphi}{\sin\theta}\partial_{
\psi}+\cot\theta\cos\varphi\partial_{\varphi}\notag\\
&\xi^L_2=\cos\varphi\partial_{\theta}+\frac{\sin\varphi}{\sin\theta}\partial_{
\psi}-\cot\theta\sin\varphi\partial_{\varphi}\notag\\
&\xi^L_3=\partial_{\varphi}
\end{align}
which are dual to the one-forms, i.e., $\langle
\xi^{R,L}_\alpha,\sigma^{R,L}_a\rangle=\delta_{\alpha a}$.
The symmetry can be explicitly shown by using the relation ${\cal
L}_{\xi^{L}_\alpha}\sigma^{R}_a=0$ 
where the ${\cal L}_{\xi_\alpha}$ is a Lie derivative along the curve generated
by the Killing vector field $\xi_\alpha$.

In the case where the rotation is present, the metric (\ref{metric3}) has less
symmetries, the $SU(2)_R$ breaks to $U(1)_R$ and the remaining Killing vectors
are given by $\xi^L_1,\xi^L_2,\xi^L_3$ and 
$\xi^R_3$.

Let us define two kinds of angular momenta
\begin{align}
L^L_{\alpha}=i\xi^L_{\alpha},\quad L^R_a=i\xi^R_a
\end{align}
which satisfy the commutation relations
\begin{eqnarray}
&&[L^L_{\alpha},L^L_{\beta}]=i\epsilon_{\alpha\beta\gamma}L^L_{\gamma},\quad
[L^R_a,L^R_b]=-i\epsilon_{abc}L^R_c,
\nonumber \\
&&\hspace{2.5cm}[L^L_{\alpha},L^R_a]=0.
\end{eqnarray}
Note that $(L^L_\alpha )^2=(L^R_a)^2\equiv L^2$ and 
the operators $L^2,L^L_3,L^R_3$ have the common eigenfunction called Wigner
$D$ function:
\begin{align}
&L^2D_{K,M}^J=J(J+1)D_{K,M}^J\\
&L^L_3D_{K,M}^J=KD_{K,M}^J\\
&L^R_3D_{K,M}^J=MD_{K,M}^J
\end{align}
where $J,K,M$ are integers satisfying $J\geq 0 $ and $K,M\leq |J|$ (we obey the
convention at \cite{Varshalovich:1988ye}).
In \cite{Murata:2007gv}, the authors demonstrated how scalar and vector fields
expand via invariant basis $\sigma^a$ 
times Wigner functions $D^J_{K,M}$. 

Let us stress that these common eigenfunctions are associated with the symmetry
generators of the rotating case. In other words, $G,K,M$ are 'good' quantum
numbers for the rotating case, where $G$ is related to $J$ but takes into
additional account the spin of the fermion. This will be discussed in more
details elsewhere. In this paper, we construct the fermionic
angular basis which preserves these quantum numbers. For sake of simplicity we
consider the static case which includes these symmetries without using
assumptions about the symmetry enhancement in this case. Our construction can
then be transposed in the more complicated case of rotating space-time with equal
angular momenta. From now on, we assume
\begin{eqnarray}
\omega(r)=0, g(r)=h(r)=r^2, b(r)=f(r)a^2(r)
\end{eqnarray}
reducing the line element to\begin{eqnarray}
ds^2&=&-f(r)a^2(r)dt^2+\frac{1}{f(r)}dr^2 \nonumber \\
&+&\frac{r^2}{4}((\sigma_1^{R,L})^2+(\sigma_2^{R,L})^2+(\sigma_3^{R,L})^2)
\label{metric_s}
\end{eqnarray}
where the coordinates have the range in $0\leq \theta<\pi$, $0\leq\varphi<2\pi$ and
$0\leq\psi< 4\pi$.

\section{Vielbein and gamma matrices}
Let us first remind the relation between the Cartesian coordinates
$(x_1,x_2,x_3,x_4)\equiv(x,y,z,w)$ and the polar coordinates
$(r,\theta,\varphi,\psi)$:
\begin{eqnarray}
&&x_1=r\sin\frac{\theta}{2}\cos\frac{\psi-\varphi}{2},~
x_2=r\sin\frac{\theta}{2}\sin\frac{\psi-\varphi}{2}\notag\\
&&x_3=r\cos\frac{\theta}{2}\cos\frac{\psi+\varphi}{2},~
x_4=r\cos\frac{\theta}{2}\sin\frac{\psi+\varphi}{2}\,.~~~~
\end{eqnarray}
Accordingly, we choose the following form for the vielbein
\begin{align}
&e_{t}^{\hat 0}=\sqrt{f}a\notag\\
&e_{r}^{\hat
1}=\frac{1}{\sqrt{f}}\sin\frac{\theta}{2}\cos\frac{\psi-\varphi}{2},~~e_{r}^{
\hat 2}=\frac{1}{\sqrt{f}}\sin\frac{\theta}{2}\sin\frac{\psi-\varphi}{2},
\notag \\
&e_{r}^{\hat
3}=\frac{1}{\sqrt{f}}\cos\frac{\theta}{2}\cos\frac{\psi+\varphi}{2},~~e_{r}^{
\hat 4}=\frac{1}{\sqrt{f}}\cos\frac{\theta}{2}\sin\frac{\psi+\varphi}{2}
\notag\\
&e_{\theta}^{\hat
1}=\frac{r}{2}\cos\frac{\theta}{2}\cos\frac{\psi-\varphi}{2},~~~~~e_{\theta}^{
\hat 2}=\frac{r}{2}\cos\frac{\theta}{2}\sin\frac{\psi-\varphi}{2},
\notag\\
&e_{\theta}^{\hat
3}=-\frac{r}{2}\sin\frac{\theta}{2}\cos\frac{\psi+\varphi}{2},~~~e_{\theta}^{
\hat 4}=-\frac{r}{2}\sin\frac{\theta}{2}\sin\frac{\psi+\varphi}{2}
\notag\\
&e_{\varphi}^{\hat
1}=\frac{r}{2}\sin\frac{\theta}{2}\sin\frac{\psi-\varphi}{2},~~~e_{\varphi}^{
\hat 2}=-\frac{r}{2}\sin\frac{\theta}{2}\cos\frac{\psi-\varphi}{2},
\notag\\
&e_{\varphi}^{\hat
3}=-\frac{r}{2}\cos\frac{\theta}{2}\sin\frac{\psi+\varphi}{2},~~~e_{\phi}^{\hat
4}=\frac{r}{2}\cos\frac{\theta}{2}\cos\frac{\psi+\varphi}{2}
\notag\\
&e_{\psi}^{\hat
1}=-\frac{r}{2}\sin\frac{\theta}{2}\sin\frac{\psi-\varphi}{2},~~~~~~e_{\psi}^{
\hat 2}=\frac{r}{2}\sin\frac{\theta}{2}\cos\frac{\psi-\varphi}{2},
\notag\\
&e_{\psi}^{\hat
3}=-\frac{r}{2}\cos\frac{\theta}{2}\sin\frac{\psi+\varphi}{2},~~~e_{\psi}^{\hat
4}=\frac{r}{2}\cos\frac{\theta}{2}\cos\frac{\psi+\varphi}{2}.
\end{align}
The vielbein satisfies the following relations:
\be
g_{MN}=e^{\hat a}_Me_{{\hat a}N}=\eta_{{\hat a}{\hat b}}e^{\hat a}_Me^{\hat
b}_N.
\ee

The form of the gamma matrices that we employ is
\begin{eqnarray}
&\gamma^{\hat{0}}=i
\begin{pmatrix}
I_2&0\\0&-I_2
\end{pmatrix}
,~\gamma^{\hat{i}}=i
\begin{pmatrix}
0&\tau^i\\-\tau^i&0
\end{pmatrix} 
,~\gamma^{\hat{4}}=
\begin{pmatrix}
0&I_2\\I_2&0
\end{pmatrix},~~~~
\end{eqnarray}
where $\tau^i~(i=1,2,3)$ are the standard Pauli matrices.  
These gamma matrices satisfy
$\{\gamma^{\hat{a}},\gamma^{\hat{b}}\}=2\eta^{\hat{a}\hat{b}}$.

\section{Dirac-Schr\"odinger equation}
\subsection{Dirac Hamiltonian}
The Lagrangian of the fermion field is 
\begin{eqnarray}
{\cal L}_{\rm fermion}= \bar{\Psi}i(\Gamma^MD_M-m)\Psi
\label{lagrangian_fermion}
\end{eqnarray}
where $m$ is the mass of the fermion. 
The gamma matrices of the curved space-time $\Gamma^M$ are defined with the help
of the vielbein $e_{\hat a}^M$ and those of the flat space-time $\gamma^{\hat a}$,ie.,
$\Gamma^M=e_{\hat a}^M\gamma^{\hat a}$. 
The covariant derivative for the fermion is defined as
\begin{eqnarray}
D_M=\partial_M+\frac{1}{8}\omega_{M\hat{a}\hat{b}}[\gamma^{\hat a},\gamma^{\hat
b}]
\end{eqnarray}
where $\omega_{M{\hat a}{\hat b}}:=\frac{1}{2}e_{\hat a}^N\nabla_Me_{{\hat b}N}$
is the spin connection. 
$M,N=0,\cdots,4$ are the 4+1 dimensional curved space-time indexes and ${\hat
a},{\hat b}=0,1,\cdots,4$
correspond to the flat tangent 4+1 Minkowski space-time.

The Dirac equation corresponding to the lagrangian (\ref{lagrangian_fermion}) is
thus
\begin{eqnarray}
\Bigl\{e_{\hat c}^M\gamma^{\hat
c}\Bigl(\partial_M+\frac{1}{8}\omega_{M\hat{a}\hat{b}}[\gamma^{\hat a},
\gamma^{\hat b}]\Bigr)-m\Bigr\}\Psi=0\,.
\label{Diraceq_form}
\end{eqnarray}

The Eq.(\ref{Diraceq_form}) can be written such as $i\partial_t\Psi={\cal
H}\Psi$.
Further, we assume that the spinor can be decomposed as 
$\Psi(t,r,\theta,\varphi,\psi)=e^{iEt}(\chi_1(r,\theta,\varphi,\psi),\chi_2(r,
\theta,\varphi,\psi))^T$. This leads to an equation of the form
\begin{equation}
{\cal H} \Psi = E\Psi,
\label{eq:diracsh}
\end{equation}
where the Dirac Hamiltonian $\cal H$ and the Dirac spinor $\Psi$ are given by 
\be
{\cal H} = \begin{pmatrix}\sqrt{f}am&-i\bar{\tau}_\mu p_\mu\\ 
i\bar{\tau}_\mu^\dagger p_\mu&-\sqrt{f}am
\label{eq:diracsh2}
\end{pmatrix}
,\ 
\Psi = \begin{pmatrix}
\chi_1 \\
\chi_2
\end{pmatrix},
\ee
%
%
where $\bar{\tau}_\mu=(-i\tau^j,I_2)~(\mu=1,\cdots,4)$ is same as one known
as the quartenion basis.
The 'momentum' is defined in terms of the coordinates and the derivatives:
\begin{eqnarray}
&&p_1=i\bar{\partial}_1+ \nonumber \\
&&
~~\frac{i}{4r}\sin\frac{\theta}{2}\cos\frac{\psi-\varphi}{2}\Bigl(a(-6\sqrt{f}
+6f+rf')+2rfa'\Bigr) 
\nonumber \\
&&i\bar{\partial}_1:=ifa\sin\frac{\theta}{2}\cos\frac{\psi-\varphi}{2}\partial_r
+\frac{2i\sqrt{f}a}{r}\cos\frac{\theta}{2}\cos\frac{\psi-\varphi}{2}
\partial_\theta 
\nonumber \\
&&~~-\frac{i\sqrt{f}a}{r}\csc\frac{\theta}{2}\sin\frac{\psi-\varphi}{2}
(\partial_\psi-\partial_\varphi)
\nonumber \\
&&p_2=i\bar{\partial}_2+
\nonumber \\
&&~~\frac{i}{4r}\sin\frac{\theta}{2}\sin\frac{\psi-\varphi}{2}\Bigl(a(-6\sqrt{f}
+6f+rf')+2rfa'\Bigr) 
\nonumber \\
&&
i\bar{\partial}_2:=ifa\sin\frac{\theta}{2}\sin\frac{\psi-\varphi}{2}\partial_r
+\frac{2i\sqrt{f}a}{r}\cos\frac{\theta}{2}\sin\frac{\psi-\varphi}{2}
\partial_\theta \nonumber \\
&&~~+\frac{i\sqrt{f}a}{r}\csc\frac{\theta}{2}\cos\frac{\psi-\varphi}{2}
(\partial_\psi-\partial_\varphi)\nonumber \\
&&p_3=i\bar{\partial}_3+
\nonumber \\
&&~~\frac{i}{4r}\cos\frac{\theta}{2}\cos\frac{\psi+\varphi}{2}\Bigl(a(-6\sqrt{f}
+6f+rf')+2rfa'\Bigr) 
\nonumber \\
&&i\bar{\partial}_3:=ifa\cos\frac{\theta}{2}\cos\frac{\psi+\varphi}{2}\partial_r
-\frac{2i\sqrt{f}a}{r}\sin\frac{\theta}{2}\cos\frac{\psi+\varphi}{2}
\partial_\theta \nonumber \\
&&~~-\frac{i\sqrt{f}a}{r}\sec\frac{\theta}{2}\sin\frac{\psi+\varphi}{2}
(\partial_\psi+\partial_\varphi)\nonumber \\
&&p_4=i\bar{\partial}_4+
\nonumber \\
&&~~\frac{i}{4r}\cos\frac{\theta}{2}\sin\frac{\psi+\varphi}{2}\Bigl(a(-6\sqrt{f}
+6f+rf')+2rfa'\Bigr) 
\nonumber \\
&&i\bar{\partial}_4:=ifa\cos\frac{\theta}{2}\sin\frac{\psi+\varphi}{2}\partial_r
-\frac{2i\sqrt{f}a}{r}\sin\frac{\theta}{2}\sin\frac{\psi+\varphi}{2}
\partial_\theta \nonumber \\
&&~~+\frac{i\sqrt{f}a}{r}\sec\frac{\theta}{2}\cos\frac{\psi+\varphi}{2}
(\partial_\psi+\partial_\varphi).
\end{eqnarray}


\subsection{Symmetries of the Dirac Hamiltonian}
The Dirac Hamiltonian does not commute with the space-time symmetry generators
$L^2,\ L_a^R,\ L_\alpha^L$. The reason is that these operators are the
generators of
the angular momenta, instead Dirac fields carry a spin, which couple to the
background angular momenta. Defining the total angular momenta:
\bea
&&G_a^R = L_a^R - \frac{1}{2}
\begin{pmatrix} 
\tau_a & 0 \\
0 & 0
\end{pmatrix}\nn\\
&&G_\alpha^L = L_\alpha^L + \frac{1}{2}
\begin{pmatrix} 
0 & 0 \\
0&\tau_\alpha
\end{pmatrix}
\eea
which  are commute with the Hamiltonian:
\be
[G_a^R,\H]=[G_\alpha^L,\H]=0.
\ee
We mention here that in the case of rotation, the commutation relations reduce
to 
\be
[G_3^R,\H^{rot}]=[G_\alpha^L,\H^{rot}]=0,
\label{test}
\ee
where $\H^{rot}$ is the Hamiltonian constructed in the rotating background.
This case will be consideration elsewhere.

As a consequence, the spinorial angular harmonics can be constructed from
eigenvectors of the $(\vec G^L)^2,\ G_3^L,\ G_3^R$. This basis would then also
be suitable for the rotating case.

\subsection{Reduction to Schr\"odinger form}


By standard procedure, one can eliminate the component $\chi_2$
from \eqref{eq:diracsh}, \eqref{eq:diracsh2}; the result is 
\begin{eqnarray}
\label{diraceigenchi1}
&&(E^2-fa^2m^2)\chi_1=
\{\bar{\tau}_\mu p_\mu
\bar{\tau}_\nu^\dagger p_\nu\chi_1\}  \nn\\
&&\hspace{0.2cm}+(E+\sqrt{f}am) \times \biggl\{\bar{\tau}_\mu
i\bar{\partial}_\mu
\frac{1}{E+\sqrt{f}am}
\biggr\}\bar{\tau}_\nu^\dagger p_\nu \chi_1.~~
\end{eqnarray}
Here the parentheses $\{~\}$ indicate the range of operation of differential in 
$\bar{\tau}_\mu p_\mu$. 
The first term in the right hand side of \eqref{diraceigenchi1} can be
computed 
as
\begin{eqnarray}
&&\bar{\tau}_\mu p_\mu
\bar{\tau}_\nu^\dagger p_\nu\chi_1
=
-a^2f^2\frac{\partial^2\chi_1}{\partial r^2} \nonumber \\
&&~~-\frac{af}{2r}\Bigl(4rfa'+3a(2f+rf')\Bigr)\frac{\partial \chi_1}{\partial r}
-\frac{4a^2f}{r^2}\Theta^2\chi_1 \nonumber \\
&&~~+\frac{a\sqrt{f}}{r^2}\Bigl(2rfa'+a(2\sqrt{f}-2f+rf')\Bigr)i{\cal D}\chi_1
-\Omega\chi_1~~~~~
\label{eq:dichi2}
\end{eqnarray}
where $\Theta, i{\cal D}$ are angular differential operators
and $\Omega$ is a function of $f,f'=df/dr$ defined by
\begin{eqnarray}
&&\Theta^2=
\csc^2\theta(\partial_\phi^2-2\cos\theta\partial_\psi\partial_\phi+\partial_\psi
^2+\sin^2\theta\partial_\theta^2\nn\\
&&\hspace{0.5cm}+\sin\theta\cos\theta\partial_\theta )
\nonumber \\
&&i{\cal D}=
i\tau_1
(\sin\psi\partial_\theta+\cos\psi\cot\theta\partial_\psi-\cos\psi\csc\theta
\partial_\phi)
 \nonumber \\
&&\hspace{0.5cm}+i\tau_2
(-\cos\psi\partial_\theta+\sin\psi\cot\theta\partial_\psi-\sin\psi
\csc\theta\partial_\phi)
\nonumber \\
&&\hspace{0.5cm}-i\tau_3\partial_\psi
\nonumber \\
&&\Omega=\frac{1}{16r^2}\Bigl\{
4r^2f^2a'^2+8raf\Bigl(-3\sqrt{f}a'+2ra'f'
\nonumber\\
&&~~~~+f(6a'+ra'')\Bigr)
+a^2\Bigl(
24f^{3/2}+12f^2-12r\sqrt{f}f'
\nonumber \\
&&~~~~+r^2f'^2
+4f(-9+9rf'+r^2f'')
\Bigr)
\Bigr\}.
\label{eq:defs}
\end{eqnarray}
The second term of the right hand side of (\ref{diraceigenchi1}) is given by
\begin{eqnarray}
&&(E+\sqrt{f}am)\times \biggl\{\bar{\tau}_\mu i\bar{\partial}_\mu
\frac{1}{E+\sqrt{f}am}
\biggr\}\bar{\tau}_\nu^\dagger p_\nu \chi_1
 \nonumber \\
&&= 
4{\cal A}rf\frac{\partial \chi_1}{\partial r}
+{\cal A}\Bigl(a(rf'-6\sqrt{f}+6f)+2rfa'\Bigr)\chi_1 \nonumber \\
&&-8{\cal A}\sqrt{f}i{\cal D}\chi_1
\end{eqnarray}
where ${\cal A}$ is defined by
\begin{eqnarray}
{\cal A}=\frac{m}{8r}\frac{af'+2fa'}{E+\sqrt{f}am}\,.
\end{eqnarray}

\subsection{Angular operator}

The operator $i{\cal D}$ can be written in terms of the matrix form
\begin{eqnarray}
i{\cal D}:=\left(
\begin{array}{cc}
-L^R_3 & \sqrt{2}L_R^+ \\
-\sqrt{2}L_R^- & L^R_3  \\
\end{array}
\right)~
\end{eqnarray}
where $L^\pm_R = (i L^R_1\pm L^R_2)/\sqrt{2}$.

Both $\Theta^2,i{\cal D}$ can be diagonalized 
by using some special combinations of the
Wigner functions $D^J_{K,M}(\theta,\psi,\phi)$, defined as
\begin{eqnarray}
&&\Theta^2 D^J_{K,M}=-J(J+1)D^J_{K,M} \\
&&L_3^R D^J_{K,M}=MD^J_{K,M} \\
&&L_R^\pm D^J_{K,M}=\pm \sqrt{\frac{J(J+1)-M(M\pm 1)}{2}}D^J_{K, M\pm 1}.
\nonumber \\
\end{eqnarray}
Note that $\Theta^2 = -(\vec L^L)^2 = -(\vec L^R)^2$ and that the Wigner
function further satisfies 
\be
L_3^L D^J_{K,M} = K  D^J_{K,M}.
\ee

The Wigner functions are the higher dimensional generalizations of the standard
spherical harmonics. The spinorial harmonics are then constructed as
follows,
\begin{eqnarray}
|0\rangle_r =
\left(
\begin{array}{c}
\sqrt{\frac{G-M}{2G}}D^{G-1/2}_{K,M+1/2} \\
\sqrt{\frac{G+M}{2G}}D^{G-1/2}_{K,M-1/2} \\
\end{array}
\right)~
\label{abasis0r}
\end{eqnarray}
satisfying
\begin{eqnarray}
&&i{\cal D}|0\rangle_r=\biggl(G-\frac{1}{2}\biggr)|0\rangle_r\\
&& \Theta^2
|0\rangle_r=-\biggl(G-\frac{1}{2}\biggr)\biggl(G+\frac{1}{2}\biggr)|0\rangle_r.
\end{eqnarray}
Similarly, a second linearly independent harmonic is given by
\begin{eqnarray}
|1\rangle_r =
\left(
\begin{array}{c}
\sqrt{\frac{G+M+1}{2G+2}}D^{G+1/2}_{K,M+1/2} \\
-\sqrt{\frac{G-M+1}{2G+2}}D^{G+1/2}_{K,M-1/2} \\
\end{array}
\right)~
\label{abasis1r}
\end{eqnarray}
which satisfies
\begin{eqnarray}
&&i{\cal D}|1\rangle_r =-\biggl(G+\frac{3}{2}\biggr)|1\rangle_r \\
&&\Theta^2|1\rangle_r
=-\biggl(G+\frac{1}{2}\biggr)\biggl(G+\frac{3}{2}\biggr)|1\rangle_r.
\end{eqnarray}
Finally we further introduce
\begin{eqnarray}
|2\rangle_r =
\left(
\begin{array}{c}
\sqrt{\frac{G+M+1/2}{2G}}D^{G}_{K,M+1/2} \\
-\sqrt{\frac{G-M+1/2}{2G}}D^{G}_{K,M-1/2} \\
\end{array}
\right)~
\label{abasis2r}
\\
|3\rangle_r =
\left(
\begin{array}{c}
-\sqrt{\frac{G-M+1/2}{2G+2}}D^{G}_{K,M+1/2} \\
-\sqrt{\frac{G+M+1/2}{2G+2}}D^{G}_{K,M-1/2} \\
\end{array}
\right).
\label{abasis3r}
\end{eqnarray}
satisfying 
\begin{eqnarray}
&&i{\cal D}|2\rangle_r
=
-(G+1)
|2\rangle_r
\\
&&i{\cal D}|3\rangle_r
=G|3\rangle_r,
\end{eqnarray}

This basis is a basis for the upper component of the Dirac spinor, satisfying 
\bea
&&G^L_3
\begin{pmatrix}
 | i \rangle_r\\
 0
\end{pmatrix}
=
K \begin{pmatrix}
 | i \rangle_r\\
 0
\end{pmatrix},\ 
G^R_3
\begin{pmatrix}
 | i \rangle_r\\
 0
\end{pmatrix}
=
M \begin{pmatrix}
 | i \rangle_r\\
 0
\end{pmatrix},\nn\\
&&(\vec G^L)^2
\begin{pmatrix}
 | i \rangle_r\\
 0
\end{pmatrix}
=
-G(G+1) \begin{pmatrix}
 | i \rangle_r\\
 0
\end{pmatrix}.
\eea

Similarly, one can define a left basis for the lower components of the Dirac
spinor:
\begin{eqnarray}
&&|0\rangle_\ell =
\left(
\begin{array}{c}
\sqrt{\frac{G+K}{2G}}D^{G-1/2}_{K-1/2,M} \\
\sqrt{\frac{G-K}{2G}}D^{G-1/2}_{K+1/2,M} \\
\end{array}
\right)~
\nonumber \\
&&|1\rangle_\ell =
\left(
\begin{array}{c}
-\sqrt{\frac{G-K+1}{2G+2}}D^{G+1/2}_{K-1/2,M} \\
\sqrt{\frac{G+K+1}{2G+2}}D^{G+1/2}_{K+1/2,M} \\
\end{array}
\right)~
\nonumber \\
&&|2\rangle_\ell =
\left(
\begin{array}{c}
-\sqrt{\frac{G-K+1/2}{2G}}D^{G}_{K-1/2,M} \\
\sqrt{\frac{G+K+1/2}{2G}}D^{G}_{K+1/2,M} \\
\end{array}
\right)~
\nonumber \\
&&|3\rangle_\ell =
\left(
\begin{array}{c}
-\sqrt{\frac{G-K+1/2}{2G+2}}D^{G}_{K-1/2,M} \\
-\sqrt{\frac{G+K+1/2}{2G+2}}D^{G}_{K+1/2,M} \\
\end{array}
\right).~
\label{angularbasis}
\end{eqnarray}

\subsection{Parity of the spinorial harmonics}

Here we discuss the parity of the basis.
In the four dimensional polar coordinates $(r,\theta,\varphi,\psi)$, the parity
transformation 
corresponds to
\begin{align}
(r,\theta,\varphi,\psi)\to(r,2\pi-\theta,\varphi+\pi,\psi-\pi).
\end{align}
The periodicity conditions for $D_{K,M}^G(\varphi,\theta,\psi)$ are (page 80,
Eq.(4) 
of \cite{Varshalovich:1988ye})
\begin{align}
&D_{K,M}^G(\varphi,-\theta,\psi)=(-1)^{K-M}D_{K,M}^G(\varphi,\theta,\psi)
\nonumber \\
&D_{K,M}^G(\varphi,\theta\pm2n\pi,\psi)=(-1)^{2nG}D_{K,M}^G(\varphi,\theta,\psi)
\nonumber \\
&D_{K,M}^G(\varphi\pm
n\pi,\theta,\psi)=(-i)^{\pm2nK}D_{K,M}^G(\varphi,\theta,\psi) \nonumber \\
&D_{K,M}^G(\varphi,\theta,\psi\pm
n\pi)=(-i)^{\pm2nM}D_{K,M}^G(\varphi,\theta,\psi)
\end{align}
where $n$ is integer.
By the parity transformation, Wigner $D$ function becomes
\begin{eqnarray}
&&D_{K,M}^G(\varphi+\pi,2\pi-\theta,\psi-\pi) \nonumber \\
&&\hspace{1cm}=(-1)^{2G+2K-2M}D_{K,M}^G(\varphi,\theta,\psi).
\end{eqnarray}
The parity of the angular basis is defined according to the parity property of
the upper component. We have four possibilities, namely $|0\rangle_r,
|1\rangle_r ,\ |2\rangle_r,\ |3\rangle_r$. The lower components are
constructed from \eqref{eq:diracsh}, \eqref{eq:diracsh2} and are respectively
proportional to
$|2\rangle_l, |3\rangle_l ,\ |0\rangle_l,\ |1\rangle_l$. It can be seen by direct inspection that parity even spinors have
$|0\rangle_r,\ |1\rangle_r$ in the upper part, while parity odd spinors have
$|2\rangle_r,\ |3\rangle_r$.

In the next section, we present a detailed basis for the Dirac spinor in the
case of the flat vacuum space-time.

\section{Radial Dirac-Schr\"odinger equation}

Now, the eigenfunction $\Psi$ 
can be separated by the angular basis and the radial part 
such as 
\begin{eqnarray}
&&\Psi^{(i)}=
\left(
\begin{array}{c}
\mathcal{F}_i(r)|i\rangle_r \\
\mathcal{G}_{i'}(r)|i'\rangle_\ell \\
\end{array}
\right),
\label{dirac_wf}
\end{eqnarray}
in which we have four possible basis $i=0,\cdots,3$. Note that for satisfying 
the full Dirac equation, special combinations $\{i,i'\}$ 
are allowed. 
It is well-known that by eliminating the lower (``the smaller'') component in (\ref{dirac_wf}), 
one can obtain the Schr\"odinger like radial equations for {\it the positive eigenvalues}  
\begin{eqnarray}
\label{radialequ} 
&&a^2f^2\mathcal{F}_i''+P_1\mathcal{F}_i'  \\
&&+\biggl(P_0^a-P_0^b( G_i-2)-\frac{a^2f}{r^2} G_i(
G_i-2)\biggr)\mathcal{F}_i=0\nonumber
\end{eqnarray}
where $G_0 = 2G+1,\ G_1 = -2G-1,\ G_2 = -2G,\ G_3 = 2G+2$ and
\begin{eqnarray}
&&P_1:=
\frac{af}{2r}\Bigl(4rfa'+3a(2f+rf')\Bigr)-4{\cal A}raf^{3/2}
 \nonumber \\
&&P_0^a:=E^2-fa^2m^2 \nonumber \\
&&\hspace{0.5cm}+\Omega-{\cal A}a\sqrt{f}\Bigl(2rfa'+a(rf'-6\sqrt{f}+6f)\Bigr)
\\
&&P_0^b:=
\frac{a\sqrt{f}}{2r^2}\Bigl(2rfa'+a(2\sqrt{f}-2f+rf')\Bigr)-4{\cal A}f; \nonumber \\
&&{\cal A}=\frac{m}{8r}\frac{af'+2fa'}{E+\sqrt{f}am}\,,
\end{eqnarray}
for $E>0$.
The lower component $\chi_2$ can be computed by using
\begin{eqnarray}
\mathcal{G}_{i'}|i'\rangle_\ell=\frac{i\tau^\dagger_{\mu}p_{\mu}}{E+a\sqrt{f}m}\mathcal{F}_i|i\rangle_r.
\label{up->dw}
\end{eqnarray}

In the similar way, {\it for the negative eigenvalues}, the equations are obtained by eliminating the 
smaller (in this case, upper) component and then
\begin{eqnarray}
\label{radialeql} 
&&a^2f^2\mathcal{G}_{i'}''+Q_1\mathcal{G}_{i'}'  \\
&&+\biggl(Q_0^a-Q_0^b( G_{i'}-2)-\frac{a^2f}{r^2} G_{i'}(
G_{i'}-2)\biggr)\mathcal{G}_{i'}=0\nonumber
\end{eqnarray}
where $G_0 = 2G+1,\ G_1 = -2G-1,\ G_2 = -2G,\ G_3 = 2G+2$ and
\begin{eqnarray}
&&Q_1:=
\frac{af}{2r}\Bigl(4rfa'+3a(2f+rf')\Bigr)-4{\cal A'}raf^{3/2}
 \nonumber \\
&&Q_0^a:=|E|^2-fa^2m^2 \nonumber \\
&&\hspace{0.5cm}+\Omega-{\cal A'}a\sqrt{f}\Bigl(2rfa'+a(rf'-6\sqrt{f}+6f)\Bigr)
\\
&&Q_0^b:=
\frac{a\sqrt{f}}{2r^2}\Bigl(2rfa'+a(2\sqrt{f}-2f+rf')\Bigr)-4{\cal A'}f; \nonumber \\
&&{\cal A}=\frac{m}{8r}\frac{af'+2fa'}{|E|+\sqrt{f}am}\,,
\end{eqnarray}
for $E<0$.
Now, the upper component can be obtained by using 
\begin{eqnarray}
\mathcal{F}_{i}|i\rangle_r=\frac{i\tau_{\mu}p_{\mu}}{|E|+a\sqrt{f}m}\mathcal{G}_{i'}|i'\rangle_\ell\,.
\label{dw->up}
\end{eqnarray}

\section{Plane waves in flat space-time}
In the case of the flat space-time, i.e., $f(r)=1$, both
Eqs.(\ref{radialequ}),(\ref{radialeql})
reduce to the standard Bessel equation:
\be
r^2
\tilde {\mathcal F}_0''+r \tilde {\mathcal F}'+\tilde {\mathcal F} \left(r^2 (E^2-m^2) -G_i^2\right)=0,
\ee
where $\tilde {\mathcal F} = r {\mathcal F}$.
The general solution is then given by
\be
\tilde {\mathcal F} = A J_{|G_i|}(kr) + B Y_{|G_i|}(k r),
\ee
where $k$ is such that $ E^2=k^2+m^2 $ and $J$ and $Y$ are the Bessel
function of first and second kind.

 $\chi_1$ has the form 
 \begin{eqnarray}
 \chi_1\sim \frac{J_{2G}(kr)}{r}|0\rangle_r.
 \end{eqnarray}
 with some multiplicative constants. 
 The lower component $\chi_2$ is estimated via 
 \begin{eqnarray}
 \chi_2=\frac{i\bar{\tau}^\dagger_\mu p_\mu }{E+m}\chi_1
 \sim\frac{J_{2G+1}(kr)}{r}|2\rangle_\ell.
 \end{eqnarray}

Here we summarize the final result of the plane wave basis set. 
The four component basis set $\{u_i\}$ which is parity even of the upper 
component is
\begin{eqnarray}
&&u^a=N_k
\left(
\begin{array}{c}
i\omega^+_{E_k}\dfrac{J_{2G}(kr)}{r}|0\rangle_r \\
\omega^-_{E_k}\dfrac{J_{2G+1}(kr)}{r}|2\rangle_\ell \\
\end{array}
\right)~\nonumber \\
&&u^b=N_k
\left(
\begin{array}{c}
i\omega^+_{E_k}\dfrac{J_{2G+2}(kr)}{r}|1\rangle_r \\
-\omega^-_{E_k}\dfrac{J_{2G+1}(kr)}{r}|3\rangle_\ell \\
\end{array}
\right).
\label{basis_u}
\end{eqnarray}
There is another four set $\{v_i\}$ with the opposite parity which is 
\begin{eqnarray}
&&v^a=N_k
\left(
\begin{array}{c}
i\omega^+_{E_k}\dfrac{J_{2G+1}(kr)}{r}|2\rangle_r \\
-\omega^-_{E_k}\dfrac{J_{2G}(kr)}{r}|0\rangle_\ell \\
\end{array}
\right)~\nonumber \\
&&v^b=N_k
\left(
\begin{array}{c}
i\omega^+_{E_k}\dfrac{J_{2G+1}(kr)}{r}|3\rangle_r \\
\omega^-_{E_k}\dfrac{J_{2G+2}(kr)}{r}|1\rangle_\ell \\
\end{array}
\right)
\label{basis_v}
\end{eqnarray}
where $\omega^+_{E_K>0},\omega^-_{E_k<0}={\rm sgn}(E_k), 
\omega^-_{E_k>0},\omega^+_{E_k<0}=k/(E_k+m)$.

If we consider a finite portion of flat space-time, the momentum $k$ is
discretized by imposing the boundary condition
\begin{eqnarray}
J_{2G+1}(k_i D)=0
\end{eqnarray}
where $D$ is a radius of a large four dimensional spherical box. The 
orthogonality condition of the basis is implemented by the integration of 
the Bessel function
\begin{eqnarray}
\int^D_0 dr r J_G(k_ir)J_G(k_jr)=\delta_{ij}\frac{D^2}{2}(J_{G+1}(k_iD))^2.
\end{eqnarray}
The nomalization constant is 
\begin{eqnarray}
N_k=\biggl[\frac{D^2}{2}(J_{2G+2}(kD))^2\biggr]^{-1/2}.
\end{eqnarray}

\section{Normal modes in $AdS$}
\subsection{Massless case}
For $AdS$ vacuum, we set $a=1,\  f=1+r^2/\ell^2 $.
In the massless case $m=0$, the asymptotic form of the solutions of 
\eqref{radialequ} or \eqref{radialeql} is
\be
              {\mathcal F}_i(r) = \frac{c_1}{r^3} + \frac{c_2}{r^2}.
\ee
Note that this result is different from the one with the setting the  limit $m \to 0$ in the
$m>0$ equation which will be seen in the next section.
 This is due to the peculiar form of the term ${\cal A}$.
The normalizability of the solutions then  requires $c_2=0$.
The expansion of the solutions around the origin leads to
\be
              {\mathcal F}_i(r) = {a_1}{r^{G_i-2}} + {a_2}{r^{-G_i}}
\ee
independently of $m$.

Let us perform the following change of variables
\be
\frac{r}{\ell} = \sqrt{\frac{1}{x^2}-1}.
\label{eq:covads}
\ee
The Schr\"odinger like equation then reduces to
\be
\left(1-x^2\right)
{\mathcal F}''-\frac{4 {\mathcal F}'}{x} +\epsilon^2 {\mathcal F} -V(x) {\mathcal F} = 0
\label{eq:m0sch}
\ee
which has to be considered for $x \in [0,1]$ and
where $\epsilon = E \ell$, and prime denotes derivative with respect to $x$. The
potential is given by
\be
        V(x) = -\frac{6}{x^2} + \frac{G_i(G_i-2)}{2(1-x)} +
\frac{(G_i+1)(G_i-1)}{2(1+x)} + \frac{1}{4} \ \ .
\ee
This equation presents singular points of second orders at $x=0, \pm 1$, but it
can be reduced to 
well-known equations after an appropriate change of the function of the form
\be
         {\mathcal F}_i(x) = x^a (1-x)^b (1+x)^c H(x).
\ee  
We find eight possible sets of the powers $a, b,c$ which reduce the order of
the singular points.
Namely, for $a=3$ and $a=2$;
we find  $b=(G_i-2)/2$ or  $b=-G_i/2$ and $c=(G_i-1)/2$ or
$c=-(G_i+1)/2$.
The  normalizability condition of the wave
function naturally
suggests $a=3$ but this leads to a Heun equation with three singular points at
$x=0,\pm 1$ which we find difficult to treat.
The choice $a=2$ leads to an hypergeometric equation for the factor $H(x)$.
Assuming $G_i >0$, the natural choice ensuring the regularity of the wave
function
at the origin  is $b=(G_i-2)/2$; then further choosing  $c=-(G_i+1)/2$ leads to
the equation
\be
       (1-x^2) H'' + (1-2 G_i - x) H' + \epsilon^2 H = 0.
\ee 
Note that the other choice of the parameter $c$ leads to a supplementary
factor $(1+x)^{G_i}$ which is regular on the domain of the interest of the variable
$x$.

Natural solutions of the equations are the Jacobi polynomials. They correspond
to the integer values
of the energy
\be
              H(x) = P^{(G_i-1, - G_i)}_n (x) \ \ , \ \ E = n .
\label{hyper}
\ee
However, these polynomials do not vanish for $x=0$ and the corresponding
wave function is non-normalizable.

In order to construct the normalizable solutions, we have to construct the
solutions of Eq.(\ref{hyper})
for arbitrary values of the energy parameter and fine-tune $E$  in such a way to
have $H(0)=0$.

For generic values of $E$, the solutions are given by hypergeometric functions
\be
           H(x) = \  _2F_1\Bigl(E,-E,G_i,\frac{1-x}{2}\Bigr).
\ee
A numerical integration of the equation leads, for fixed $G_i$, to a family of
solutions which fulfill 
the condition $H(0)=0$. As can be expected, the solutions are labeled by the
number of zeros in $[0,1]$.

\begin{figure}[h]
\begin{center}
\includegraphics[width=9cm]{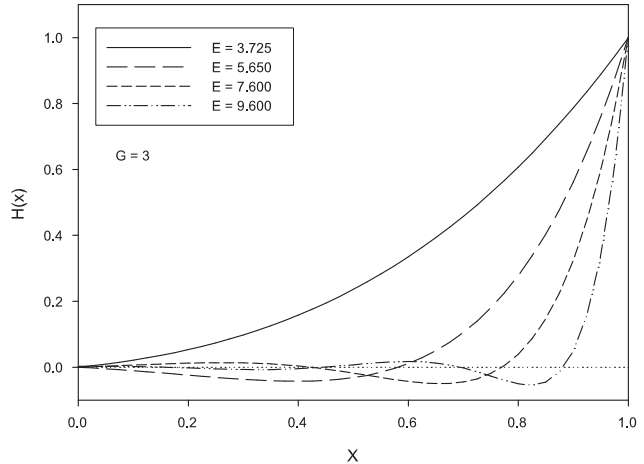}
\end{center}
\caption{Profile of the first four functions $H(x)$ for $G_0=3$. We have set
$\ell=1$.
}
\label{fig:hyper}
\end{figure}

Setting $G_0=3$, the profiles of the four first functions are presented in Fig.
\ref{fig:hyper}. The normalization
$H(1)=1$ is chosen. 
For $G_0=3,4$, we find respectively for the four first energy levels
\be
        \epsilon = 3.725, 5.65, 7.62 , 9.6 \ \ , \ \ \ \epsilon = 4.76 , 6.68,
8.64 , 10.62.
\ee

Let us stress that the spectrum is symmetric for $\epsilon\rightarrow
-\epsilon$, as can already be seen in Eq.\eqref{eq:m0sch}.

We believe there is a direct connection with the results of
\cite{Camporesi:1995fb}. Indeed, we find a general solution in terms of the
hypergeometric function of half variable. In the case of
\cite{Camporesi:1995fb}, the spectrum was continuous because the spinor lived
on a compact space. Here the space is not compact and we need to impose a
normalizability condition.

Note that we indirectly recover the results of \cite{LopezOrtega:2007sr} for
deSitter space. Indeed, the condition that the solution is regular at its poles
implies that the eigenvalue is proportional to an integer. Here, this case gives
non-normalizable eigenstates.

\subsection{Massive case}
The equations in $AdS$ vacuum are invariant under the rescaling
$m' = m\ell,\ E' = E\ell$ and $r' = r/\ell$, we therefore set arbitrarily
$\ell=1$ without loss of generality.

The asymptotic solution to \eqref{radialequ} is given by
\begin{equation}
{\mathcal F} \approx {\mathcal F}_\infty = A_\infty r^{-2-m } + B_\infty r^{-2+m }.
\label{eq:asads}
\end{equation}
The spinor is normalizable if it decays at least as $r^{-3}$, implying that no
normalizable mode exists for $|m|< 1$. 

It is important to remark that as mentioned in the previous section, the
asymptotic behavior of the massless case differs from the $m=0$ limit of the
massive case. Thanks to this fact, normalizable modes exist in the massless
case.
Close to the origin, the function $\mathcal{F}$ behaves like
\begin{equation}
{\mathcal F}\approx A_0 r^{G_i-2} + B_0 r^{-G_i }.
\label{eq:asogn}
\end{equation}

\begin{figure*}[t]
  \begin{center}
\includegraphics[scale=1.5]{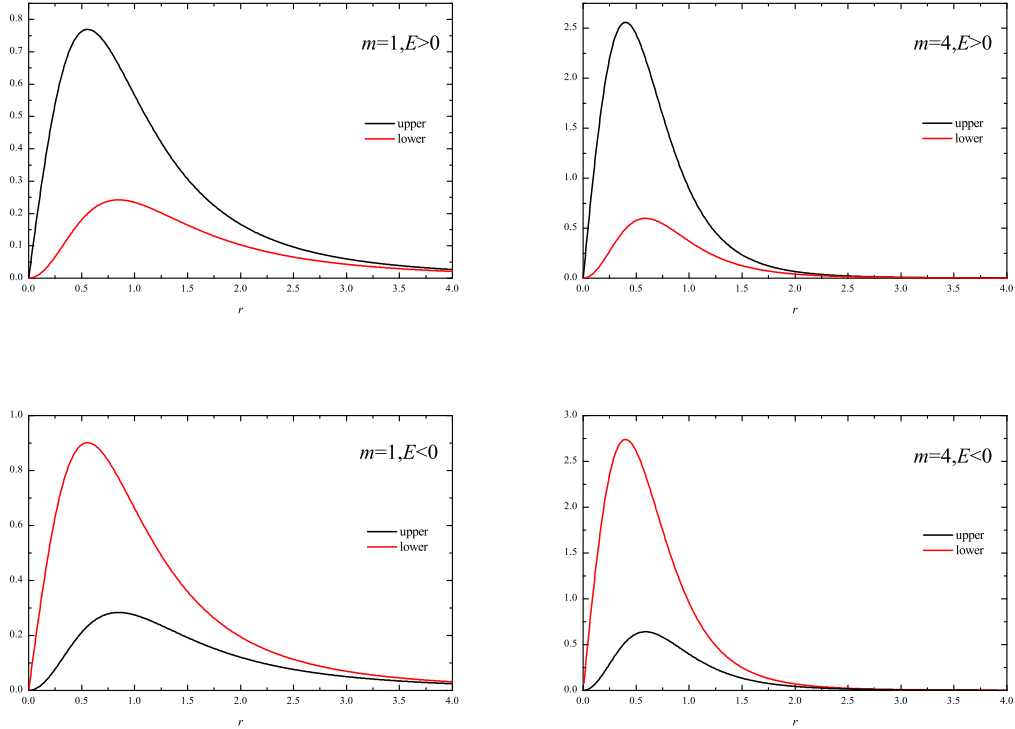}
  \caption{The ground state of the spinor for the case of mass parameter $m=1,4$ 
of $\mathcal{F}_0$ (the top two figures) and $\mathcal{G}_0$ (the bottom two figures).
For larger value of $m$, the small components decrease. }
  \label{diracspinor}
  \end{center}
\end{figure*}

In order to construct the spectrum of the equation for $m > 0$,
it is useful to use the change of variable \eqref{eq:covads} and to perform the
following
change 
of function on the radial function ${\mathcal F}_i(r)$ appearing in Eq. \eqref{radialequ}:
\be
{\mathcal F}_i(x) = x^{m+2} \ (1-x)^{(G_i-2)/2} \ (1+x)^{(G_i-1)/2} H(x).
\ee
If there exists solutions $H(x)$ which are regular at $x=0$ and $x=1$, 
the factorization of the power of $x$ and of the power of $(1-x)$ 
ensure respectively the normalizability (i.e. for $r\to \infty)$ and the 
regularity (i.e. for $r=0$)
of the corresponding wave function ${\mathcal F}_i(r)$.

The equation for the new function $H(x)$ is then found to be
\be
x(1-x^2)(E x + m) \frac{d^2 H}{dx^2} + P_3(x) \frac{d H}{dx} + P_2(x) H =0,
\label{four_p}
\ee
where
\bea
&& P_3(x) = - E(2G_i+2m+1) x^3 \nn\\
&&\hspace{0.5cm}+ (E-2 G_im - 2 m^2 - 2m) x^2 \nn\\
&&\hspace{0.5cm}+ m(2E+1)x + m(2m+1),
\eea
\bea
     &&P_2(x) = -(G_i+m-E) (E(E+m+G_i) x^2 \nn\\
     &&\hspace{0.5cm}+ m(E + m + G_i +1)x - m).
\eea
It turns out that Eq. (\ref{four_p}) admits a family of solutions which are
polynomials of
degree $n$ in $x$, say $H(x) = H_n(x)$.
The energy $E$ of the solution $H_n(x)$ is given by
\be
             E_n = (-1)^n (G_i+m+n).
\label{energy}
\ee
Note stress that only even $n$~(i.e., $E_n>0$) in (\ref{energy})
are allowed because 
the Eq.(\ref{radialequ}) is of the positive eigenstates. 
To get the solutions of the negative eigenstates, 
we should start with  (\ref{radialeql}).  
We have the same form as (\ref{energy}), but at that case 
we should adopt only for the odd $n$. 

The generic coefficients of the polynomials $H_n(x)$ can be constructed  by
solving some recurrence relations.
The even and odd parts of the polynomials split naturally, 
leading to a form
 \bea
&& H_n(x) = H_{n,e}(x) +  H_{n,o}(x) \  ,\\
&&\ H_{n,e}(-x) = H_{n,e}(x)
\ ,
\ H_{n,o}(-x) = - H_{n,o}(x).\nn
\eea
Assuming first that the degree of the polynomial $H_n$ is even, i.e. $n=2p$,
then we have
\be
      H_{n,e}(x) = \sum_{j=0}^p	 (-1)^j C_j^p a^{ee}_jx^{2j}
 \ \
,
\ee
\be
      H_{n,o}(x) = n \sum_{j=0}^{p-1} (-1)^j C_j^{p-1} a^{eo}_jx^{2j+1}
\ee
where $C_j^p$ denotes the combinatoric symbols. The ratios of double factorials
are in fact
 finite products.
In the above expressions, we have defined for compactness
\bea
&&a^{ee}_j=\frac{(2m+2p-1)!!}{(2m+2j-1)!!}\frac{(2G_i+2m+2p+2j-1)!!}{
(2G_i+2m+4p-1)!!},\nn\\
&&a^{eo}_j=\frac{(2m+2p-1)!!}{(2m+2j+1)!!}\frac{(2G_i+2m+2p+2j-1)!!}{
(2G_i+2m+4p-1)!!}.\nn
\eea
 
For  odd values of $n$, i.e. $n = 2p+1$ the polynomials have similar form~:
\be
      H_{n,e}(x) = \sum_{j=0}^p	 (-1)^j C_j^p a^{oe}_jx^{2j} \ \
,
\ee
\be
      H_{n,o}(x) = -  \sum_{j=0}^{p} (-1)^j C_j^{p} a^{oo}_j x^{2j+1}
\ee
where
\bea
a^{oe}_j
=\frac{(2m+2p+1)!!}{(2m+2j-1)!!}\frac{(2G_i+2m+2p+2j+1)!!}{(2G_i+2m+4p-1)!!},
\nn\\
a^{oo}_j = \frac{(2m+2p+1)!!}{(2m+2j+1)!!}
\frac{(2G_i+2m+2p+2j+1)!!}{(2G_i+2m+4p+1)!!}.\nn
\eea

It is possible to express these series in term of a combination of Jacobi
polynomials, defined as
\bea
&&P_p^{\alpha,\beta}(z) =
p!\frac{\Gamma(\alpha+p+1)}{\Gamma(\alpha+\beta+p+1)}\nn\\
&&\hspace{0.5cm}\times\sum_{j=0}^p
C^p_j\frac{\Gamma(\alpha+\beta+p+j+1)}{\Gamma(\alpha+j+1)}
(\frac{z-1}{2})^j.
\eea

Indeed, noting that for a constant $N$, 
\be
N !! = 2^{N/2} \Gamma\left(\frac{N}{2} +1\right),
\ee
the coefficients of the series defining our solutions turn out to be precisely
those of the Jacobi polynomials. 

For even degree polynomials, we find 
\bea
&&H_{2p}(x) = 
\frac{1}{C^{m+G+p-\frac{1}{2}}_p} \nn \\
&&\times
\left( P_p^{m-\frac{1}{2},G_i}(1-2x^2)
+ x P_{p-1}^{m+\frac{1}{2},G_i}(1-2x^2) \right)~~~~
\label{sol_even}
\eea
with the associates eigenvalue given by $E_{2p} = G_i+m+2p$.

Similarly, for odd degree polynomials, we find
\bea
&&H_{2p+1}(x)
= \frac{m+p+\frac{1}{2}}{(p+1)C_{p+1}^{m+G+2p+\frac{1}{2}}}P_p^{m-\frac{1}{2},
G_i+1}(1-2x^2) \nn\\
&&\hspace{0.5cm}- \frac{x}{C_p^{m+G+2p+\frac{1}{2}}}
P_p^{m+\frac{1}{2},G_i}(1-2x^2).
\eea

The lower component of the Dirac spinor formally 
can be reconstructed by using (\ref{up->dw}), but it is not 
straightforward and is tedius task to get the explicit form 
because the solution (\ref{sol_even}) is already quite complicated.  
The massive spectrum we build here agrees with the results of
\cite{Cotaescu:2007xv}.

\begin{figure}[t]
  \begin{center}
\includegraphics[scale=.85]{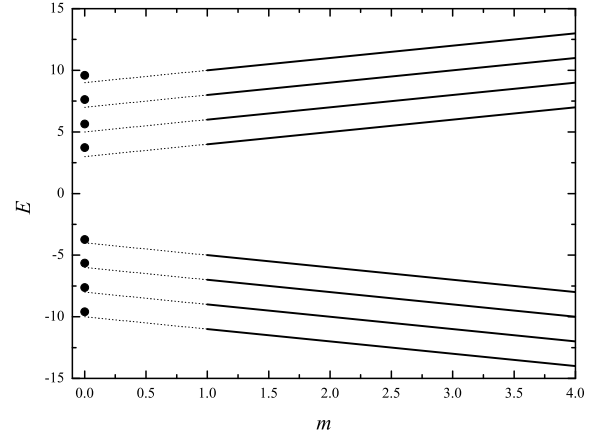}
  \caption{The eigenvalues as a function of the mass parameter $m$. The dots on $m=0$ denote 
the result of the massless case.}
  \label{mass}
  \end{center}
\end{figure}

\subsection{Numerical study}
We have found analytically the  normalizable solutions for the Schr\"odinger like 
equations (\ref{radialequ}), (\ref{radialeql}) for both the massless/massive cases.
If one obtains the smaller component of the solutions, formally (\ref{up->dw}) and (\ref{dw->up}) should be used.
It is, however, tedius task to compute analytically.  
In this section we numerically solve the equations and obtain both the larger and the smaller components. 

It is convenient to use the rescaled coordinate defined as $y:=r/(1+r)$ which runs from 0 to 1. 
The equations can be written by the new coordinate
\begin{eqnarray}
&&\bigl\{(1-y)^2+y^2\bigr\}^2\frac{d^2\mathcal{F}_i}{dy^2}\notag\\
&&+\biggl[p_{1}(1-y)^2-2\frac{\bigl\{(1-y)^2+y^2\bigr\}^2}{1-y}\biggr]\frac{d\mathcal{F}_i}{dy}\notag\\
&&+\biggl[p_{0}^a-p_{0}^b( G_i-2)-\frac{(1-y)^2+y^2}{y^2}G_i(G_i-2)\biggr]\mathcal{F}_i
=0 \nonumber \\
\end{eqnarray}
where
\begin{eqnarray}
&&p_1:=\frac{3\bigl\{(1-y)^2+y^2\bigr\}\bigl\{(1-y)^2+2y^2\bigr\}}{y(1-y)^3}\nonumber\\
&&\hspace{1.5cm}-4\frac{\mathcal{A}y\bigl\{(1-y)^2+y^2\bigr\}^{3/2}}{(1-y)^4}\nonumber\\
&&p_{0}^a:=|E|^2-m^2+\frac{m^2y^2}{(1-y)^2}+\Omega 
-\frac{{\cal A}\sqrt{(1-y)^2+y^2}}{(1-y)^3}\nonumber \\
&&\hspace{1.0cm}\times
\Bigl(6(1-y)^2+8y^2-6(1-y)\sqrt{(1-y)^2+y^2}\Bigr)\nonumber\\
&&p_{0}^b:=\frac{\sqrt{(1-y)^2+y^2}}{y^2}\Bigl(-(1-y)+\sqrt{(1-y)^2+y^2}\Bigr) \nonumber \\
&&\hspace{1.5cm}-4\frac{\mathcal{A}(1-y)^2+y^2}{(1-y)^2}
\end{eqnarray}
\begin{eqnarray}
&&{\cal A}=\frac{m}{4}\frac{1-y}{E(1-y)+m\sqrt{(1-y)^2+y^2}}\nonumber\\
&&\Omega=\frac{1}{4y^2}\biggl\{
6(1-y)\sqrt{(1-y)^2+y^2} \nonumber \\
&&
\hspace{1cm}-6(1-y)^2+17y^2+\frac{24y^4}{(1-y)^2}
\biggr\}.
\end{eqnarray}
The asymptotic behaviors (\ref{eq:asads}),(\ref{eq:asogn})
now become
\begin{align}
\mathcal{F}_{\infty}=A_{\infty}(y-1)^{2+m}+B_{\infty}(y-1)^{2-m}
\end{align}
for the AdS boundary and 
\begin{align}
\mathcal{F}=A_0y^{G_i-2}+B_0y^{-G_i}
\end{align}
near to the origin.

For the numerical analysis, we employ a scheme based on simple first order 
perturbation, which is quite efficient for the present 
eigenvalue problem (see, for example, \cite{Brihaye:2010nf} and also the 
reference in the paper). 
The scheme is summarized as follows.
\begin{enumerate}
\item[(i)]
We assume an eigenvalue $E_0$, and solve the equation for $\mathcal{F}_i(y)$ from $y=0$ to an intermediate 
point $y=y_{\rm fit}$ by using the standard Runge Kutta method.
\item[(ii)]
The value of the solution $\mathcal{F}_i(y)$ at $y_{\rm fit}$ fits with the asymptotic solution $\mathcal{F}_{\infty}$
by a multiplicating factor $\alpha$
\begin{eqnarray}
\alpha \mathcal{F}_{\infty}(y_{\rm fit})\equiv \mathcal{F}_i(y_{\rm fit}).
\end{eqnarray}
\item[(iii)] 
We introduce 
a $\delta$-functional potential defined by
\be
V_\delta(y):=-\frac{[\mathcal{F}_i'(y_{\rm fit})]^{y_{\rm fit}+0}_{y_{\rm fit}-0}}{\mathcal{F}_i(y_{\rm fit})}~\delta (y-y_{\rm fit}).
\ee
If such $\delta-$functional potential exists, the eigenfunction is still continuous at $y_{\rm fit}$ but the derivative is not. 
Correction in terms of the first order perturbation 
\begin{eqnarray}
&&\Delta E= \int \frac{y^3dy}{(1-y)^5}
\mathcal{F}_i^*(y)V_\delta(y)\mathcal{F}_i(y) \nonumber \\
&&=-\frac{y_{\rm fit}^3}{(1-y_{\rm fit})^5}[\mathcal{F}_i'(y_{\rm fit})]^{y_{\rm fit}+0}_{y_{\rm fit}-0}\mathcal{F}_i(y_{\rm fit})
\end{eqnarray}
efficiently improves the eigenvalue, i.e., the eigenfunction.
\item[(iv)]
The process (i)-(iii) is repeated until the convergence is attained. 
If the analysis reaches the correct eigenfunction, it no longer has discontinuity 
and the computation is successfully terminated. 
\end{enumerate}
We present some typical results of the eigenvalues in Table \ref{eigenvalues}.

\begin{table}[htbp]
\begin{center}
\begin{tabular}{c|ccc}
	\hline
	\hline
   & $n=0$ & $n=1$       &$n=2$  \\
	\hline	
     $\mathcal{F}_i$	   &  4.00000003 &  6.00000008  &  8.00000015  \\
$\mathcal{G}_i$& -5.00000024 & -7.00000069 & -9.00000107 \\
	\hline
	\hline
\end{tabular}
\caption{\label{eigenvalues}The eigenvalues for $l=1,m=1,G_i=3$.}
\end{center}
\end{table}

Now we numerically calculate the lower component of $\mathcal{F}_i$ and also the upper component of $\mathcal{G}_i$.
In Fig.\ref{diracspinor}, we plot the $\mathcal{F}_0, \mathcal{G}_0$ and the corresponding lower component for the case of 
$m=1$ and $m=4$. Ratio of the lower component to the upper component is smaller for $m=4$. Thus,  
we speculate that for $m\to\infty$, only upper component remains, corresponding to the ``non-relativistic limit''.
In Fig. \ref{mass}, we show the eigenvalues for the case of the normalizable mode.

\section{Gauss-Bonnet gravity and boson star}
The above discussion can be extended in numerous directions.
 Namely~: (i) the gravity sector  can be extended
by a Gauss-Bonnet term, (ii) various matter fields can be supplemented to the Einstein or Einstein-Gauss-Bonnet (EGB) action.
To be more concrete, the Dirac equation can be studied in the background of a space-time constructed out of the model
\be
\label{egbbs}
   S = \frac{1}{16 \pi G} \int d^5 x ( R - 2 \Lambda + \frac{\alpha}{2} L_{GB} 
   - (16 \pi G) \partial_M \Pi^{\dagger} \partial^M \Pi)              
\ee
where $R$ is the Ricci scalar, $\Lambda=-6/\ell^2$ is the cosmological constant,
  $\Pi$ is a complex field,  $\alpha$ denotes the Gauss-Bonnet coupling constant
and $L_{GB}$ is the Gauss-Bonnet term, constructed out of the Riemann tensor in the standard way~:
\be
        L_{GB} = R^{MNKL} R_{MNKL} - 4 R^{MN} R_{MN} + R^2 \ \ , 
\ee
with $M,N,K,L \in \{0,1,2,3,4 \}$.
In the case of  EGB gravity, the usual $AdS$ space-time is modified by the Gauss-Bonnet interaction, the metric
function $f(r)$ takes the form~:
\be
\label{egb_ads}
        f(r) = 1 + \frac{r^2}{\ell_c^2} \ \ ,  \ \
        \frac{1}{\ell_c^2} = \frac{1}{\alpha}\biggl[1 - \sqrt{1 - \frac{2\alpha}{\ell^2}}\biggr]
\ee
In the presence of the cosmological constant, the  range
of the Gauss-Bonnet coupling constant is limited: $\alpha \in [0, \frac{\ell^2}{2}]$, the upper limit
is called the Chern-Simons limit \cite{chamseddine}. 

The model (\ref{egbbs}) is one of the simplest way to couple matter (minimally) to gravity.
The coupled system admits regular, localized, stationnary solutions called  boson stars 
(see e.g. \cite{Mielke:1997re} for a review). 
For $d > 4$ boson stars were constructed namely \cite{Astefanesei:2003qy,Prikas:2004fx,Hartmann:2013tca}.  
 Even in the absence of a self-interacting potential of the scalar field, the asymptotically $AdS$ space-time
 renders possible the existence boson stars.
 
 The simplest boson star can be constructed by performing a spherically symmetric ansatz for the scalar field~:
  ~: $\Pi(x) = \exp(-i \omega t) \phi(r)$, 
  where the time-dependant harmonic factor contains the 
   frequency parameter $\omega$ and $r$ is the radial variable. With this ansatz and the metric (\ref{metric}),
   the field equations reduces to a system of three differential equations for the functions $f,b, \phi$.
   These equations have to be solved numerically. However, in the limit where the scalar field decouples from gravity
   (the so called probe limit) the gravity part is determined by (\ref{egb_ads}) while the 
    the Klein-Gordon equation
 of the massless boson takes the form
 \be
      (r^3 f \phi')' - \frac{r^3 \omega^2}{f} \phi = 0 
 \ee
 which can be solved in terms of hypergeometric functions
 \be
      \phi(r)  = \frac{\ell_c^4}{(r^2 + \ell_c^2)^2} 
     \phantom{F}_2 F_1 \biggl(\frac{4-\omega \ell_c}{2},\frac{4+\omega \ell_c}{2};3, \frac{\ell_c^2}{r^2 + \ell_c^2}\biggr) \ .
 \ee
The regularity of this solution on the full line requires $\omega \ell_c = 4 + 2 k$ with an integer $k$. 
These corresponding solutions,  called oscillons \cite{Cardoso:2004nk},
are  regular,  localized in  space and labelled  by the integer  $k$ 
which sets the number of nodes of the profile $\phi(r)$ and  the frequence $\omega$). 

Once coupled to gravity, the oscillons solutions get deformed by the geomety 
and exist on a finite range of the frequency $\omega$; from now on, we pose $\kappa \equiv 16 \pi G$. 
The  three coupled differential equations for the functions $f,b,\phi$ 
have to be solved with appropriate boundary conditions. At the center of the boson star, the regularity of the
solutions requires $f(0)=1$, $b'(0)=0$, $\phi'(0)=0$. 
 Expanding the fields about the origin leads to the following behaviour
\bea
    f(r) &=& 1 + F_2 r^2 + O(r^4) \ , \nonumber \\   
    b(r) &=& B_0 + \frac{6 B_0 F_2 - \kappa \omega^2 \Pi_0^2 - 12 B_0/\ell^2}{6(\alpha F_2 - 1)} r^2 + O(r^4) \ ,
    \nonumber \\ 
    \phi(r) &=& \Pi_0 - \frac{\Pi_0 \omega^2}{8 B_0} r^2 +  O(r^4)
\eea 
where the parameters $B_0$, $\Pi_0$ are undertermined  while
\be
\label{f_2_value}
          F_2 = \frac{6 B_0 - \sqrt{36 B_0^2 + 6\alpha B_0 \kappa \omega^2 \Pi_0^2 - 72 \alpha B_0^2/\ell^2 }}{6 \alpha B_0}.
\ee

Asymptotically, the scalar field is required to vanish 
while the metric (\ref{egb_ads}) is approached. More precisely, for $\alpha < \ell^2/2$, the fields should obey
 \bea
 \label{asymptotic}
        f(r) &=& 1 + \frac{r^2}{\ell_c^2} + \frac{\cal \tilde M}{r^2} + O(r^{-4}) \ \ , \nonumber \\ 
       b(r) &=& 1 + \frac{r^2}{\ell_c^2} + \frac{ \cal M}{r^2} + O(r^{-4}) \ \ , \nonumber \\
        \phi(r) &=& \frac{\Pi_{\infty}}{r^4} + O(r^{-6}).     
 \eea 
 (the limit $\alpha = \ell^2/2$ is special, see e.g. \cite{Brihaye:2013vsa}).
Technically, the equations are integrated numerically
by a fine tuning the parameters $\omega$, $B_0$ and $\Pi_0$ 
in such a way that the boundary conditions are obeyed. The Newton constant $\kappa$
can be rescaled in the scalar field and the Gauss-Bonnet parameter is fixed by hand.
For simplicity, we address only the deformations of the fundamental oscillons characterized by $k=0$.
With a choice of $\alpha$, a branch of boson stars labelled by the frequency $\omega$ 
(or equivalently by the value $\phi(0) = \Pi_0$)
can be constructed. 
The boson stas are namely   characterized by  mass~:
\be
       M = \frac{V_3}{16 \pi G} ({ \cal \tilde M} - 4{\cal M}) (1 - \frac{2\alpha}{\ell^2})^{1/2} \ \ , \ \
       V_3 = 2 \pi^2 
\ee
where the parameters ${\cal M}, {\cal \tilde M}$ appear in the asymptotic of the metric (\ref{asymptotic}).
We construct numerically a large number of solutions which allow an understanding of the pattern of the EGB boson stars. 
For all values parameter $\phi(0)$ that we have considered, the EGB boson star could 
be constructed up to the maximal value $\alpha = 1/2$;  it was checked that the expression under the square
root in (\ref{f_2_value}) stays strictly positive. (Potentially this term can become negative
for some values of $\alpha$, limiting the domain of existence of the solutions.
For instance, this happens for spinning solitons \cite{Henderson:2014dwa}). 

The dependence of the mass of the fundamental (i.e. with $k=0$) boson stars on the frequency  $\omega$ is reported on Fig. \ref{for_terence} for
several  values of $\alpha$ (the mass is in the unit $V_3/ 16 \pi G$).
 
\begin{figure}[h]
\begin{center}
{\label{for_terence}\includegraphics[width=8cm]{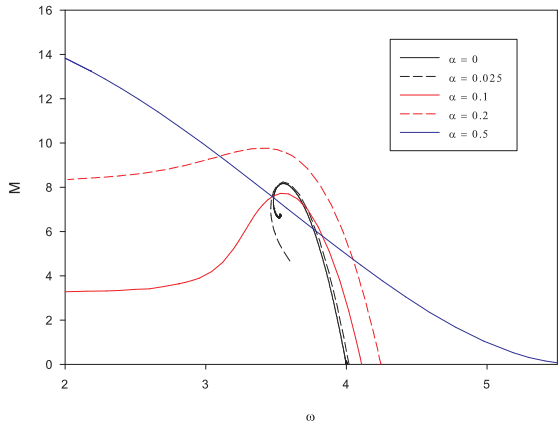}}
\end{center}
\caption{Mass dependence on the frequency for boson stars for four different values of $\alpha$.
\label{for_terence}
}
\end{figure}
In the case of pure Einstein gravity, the 
curve is represented by the solid-black line.  
In the limit $\omega \to 4$, the scalar field tends uniformly to zero and the $AdS$ space-time is approached
uniformly. Increasing gradually the central value $\phi(0)$ reveals that the mass reaches a maximum
and that the boson star exists only on a finite interval of frequencies, $\omega \in [3.5, 4.0]$.
In that limit $\phi(0) \to \infty$, the metric seems to approach a configuration with a singularity
of the Ricci scalar at the origin.
The mass remains finite and the $M-\omega$ line takes the form of a spiral.

When the Gauss-Bonnet parameter is non-zero, the solution  is gradually deformed. In particular
the curve $M-\omega$ changes smoothly,
but the spirals have a tendency to disappear. This feature seems to be generic
for EGB boson stars;
it is first observed in \cite{Hartmann:2013tca} in the case of asymptotically flat space-time
(and in the presence of a self-interacting potential of the scalar field)
and more recently in \cite{Henderson:2014dwa,Brihaye:2013zha} for 
asymptotically $AdS$ spinning boson and non-spinning solutions.
When the parameter $\alpha$ becomes large enough, our numerical analysis strongly indicates that
the interval of allowed frequencies is extended to $[0, 4/ \ell_c]$ 
(in particular $\omega \in [0, 4 \sqrt{2}]$  for the Chern-Simons limit 
$\alpha = 1/2$). A numerical evaluation
of the critical value of $\alpha$ is quite involved and is not aimed for this paper.

%

One question which occurs naturally is the study of the evolution of the spectrum of the fermion
in the EGB space-time and/or in the space-time of a boson star. 
In order to study the effect of the scalar field on the fermion spectrum, we
introduce a Yukawa type coupling, by extending the fermionic Lagrangian
\eqref{lagrangian_fermion} to

\begin{eqnarray}
{\cal L}_{\rm fermion}= \bar{\Psi}(i\Gamma^MD_M-m - \mu |\phi|)\Psi,
\label{lagrangian_fermion_yuk}
\end{eqnarray}
where $\mu$ is the Yukawa coupling and $|\phi|$ is the norm of the scalar
field. The Dirac and Dirac-Schr\"odinger equations are extended
straightforwardly, according to the procedure described in Sec.IV. 
All the symmetry consideration of that section
remain unaltered, in particular the angular sector is unaffected.

In the absence of matter field and for a massless fermion the fermionic levels obey the scaling
rule $E(\alpha) = E(0)/\ell_c(\alpha)$. As a consequence, the energy levels
increase with the increasing $\alpha$.
In the presence of a fermion mass, the scaling rule is only approximative, but the increase of the fermion 
levels with $\alpha$ still holds. 

We finally study the response of the fermion eigenmodes to boson stars  in EGB space-time
by considering the reduced Dirac equation in the background of an oscillon. The sketch of these
results  are summarized in Fig. \ref{probe_bs} for the first two fermionic modes for Einstein and EGB oscillons. 
We check that the qualitative properties of the fermionic eigenvalues in the case of the 
boson stars (i.e. for scalar minimally coupled to gravity)  are qualitatively the same.
\begin{figure}[h]
\begin{center}
{\label{probe_bs}\includegraphics[width=8cm]{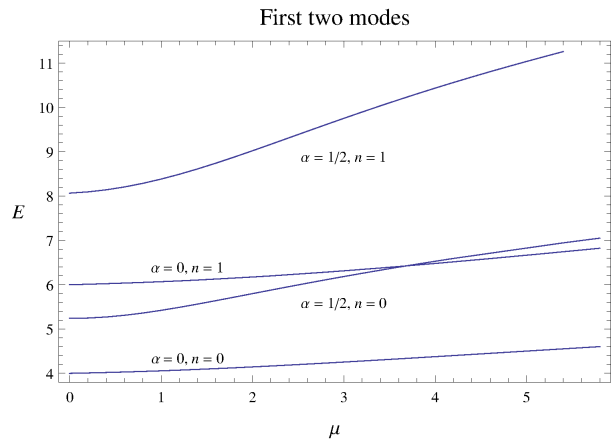}}
\end{center}
\caption{Dependence of the fundamental fermionic modes and of the first excited mode 
in the background of an oscillon as function of the Yukawa coupling constant $\mu$ 
\label{probe_bs}
}
\end{figure}
In the case of Einstein boson star, the fermionic levels, obtained numerically, increase
slightly with the Yukawa parameter, indicating that the fermion becomes more strongly bounded to the space
of the boson star. 
As expected, in the case of the boson stars in the EGB-gravity (and then also in the Chern-Simons limit)
the fermion binding energy becomes  much stronger. 
This result can be explained by the stronger 'harmonic oscillator' coupling played by the underlying
 space-time asymptotically.


\section{Conclusion}
In this paper, we discuss in details the separation of the radial and angular
variables for the Dirac equation in a 5-dimensional space-time. The full
isometry group of the angular sector of the metric is $SU(2)_R\times
SU(2)_L$ symmetry. However, only the subgroup
$SU(2)_R\times U(1)_L$ appears as a manifest symmetry in the coordinates
used here. This subgroup is precisely the isometry group of the 5-dimensional
 Myers-Perry space-time with equal angular momentum. The angular basis
obtained here can be used for the case of rotating space-time with
equal angular momenta. 
As a crosscheck of our equations we (re)derived the plane wave basis in flat
vacuum space-time and the normal basis for the case of
$AdS$ vacuum in both massless and massive case.

The massless spinor in $AdS$ is expressed in terms of the hypergeometric functions,
and the spectrum is symmetric under change of sign of the eigenenergy. 

The construction of the eigenvector in this case cannot be done till the end
algebraically, and the corresponding eigenvalue has to be be determined
numerically by imposing the suitable boundary condition.

In contrast, the massive case can be treated in terms of Jacobi polynomials and
lead to a set of explicit eigenvalues \cite{LopezOrtega:2007sr}. Accordingly,
the spectrum in the
massless case doesn't exhibit the same amount of regularity
as for the massive case. Furthermore, they are not continuously connected. This
constitutes one of our original result.

The massive eigenvalues can be labeled by an integer $n$ denoting the degree of
the polynomial $H(x)$ appearing in the construction. They are alternatively
positive and negative as $n$ increases, and the associated eigenfunction is
expressed as a peculiar linear combination of two Jacobi polynomials. Positive
eigenvalues are associated to even degree polynomials,
while the negative energy modes have odd degree.

We note that the solution we describe here are analytic on the whole $AdS$
space-time since both normalizability and regularity at the origin are imposed.

Our results are relevant for many purposes. For instance, they
be used as a basis of the Dirac spinor in asymptotically $AdS$ and flat
space-time, which can be used to construct the spectrum of the Dirac
equation in more elaborate setup, e.g. fermionic modes around a holographic
superconductor. 

The simplest model discribing an such a physical system consists of a charged, complex scalar field
minimally coupled to gravity. The relevant classicals solution is determined numerically. In Sect VIII
we made a step forward in this direction by studying the evolution of the Dirac equation in the background
of the space-time of a boson star supported by Einstein and Einstein-Gauss-Bonnet gravity. Here the eigenvalues
are determined numerically and our results show that the spectrum is smoothly deformed by the boson star. 

Finally, let us stress again that the symmetry preserved in the construction is
the isometry group of the rotating case (with equal angular momenta). This is
the main result of this technical paper.
This works settles the issue of the angular part for this case. Of course, more
work is
needed and is in progress, but will be presented elsewhere due to the various
technical points involved in the full construction.

\section*{Acknowledgement}
We thank Jorge Rocha for useful discussion and comments. 
N.S. and H.Y. appreciate the kind consideration and assistance 
at Universit\'{e} de Mons for our several times of visit and stay. 
H.Y. acknowledges the financial support of Tokyo University of Science.
T.D. also acknowledges the Tokyo University of Science for kind hospitality during the
last stages of this project. 
The work of Y.B. and T.D. was supported in part by an ARC contract n° AUWB-2010/15-UMONS-1.

\appendix

\bibliographystyle{utphys}
\bibliography{fermion_superconductor}

\end{document}